\documentclass{article}

\usepackage[T1]{fontenc}

\def\fa{\forall}
\def\ra{\rightarrow}
\def\la{\leftarrow}

\newbox\tempa
\newbox\tempb
\newdimen\tempc
\def\mud#1{\hfil $\displaystyle{\mathstrut #1}$\hfil}
\def\rig#1{\hfil $\displaystyle{#1}$}
\def\irulehelp#1#2#3{\setbox\tempa=\hbox{$\displaystyle{\mathstrut #2}$}%
		        \setbox\tempb=\vbox{\halign{##\cr
	\mud{#1}\cr
	\noalign{\vskip\the\lineskip}%
	\noalign{\hrule height 0pt}%
	\rig{\vbox to 0pt{\vss\hbox to 0pt{${\; #3}$\hss}\vss}}\cr
	\noalign{\hrule}%
	\noalign{\vskip\the\lineskip}%
	\mud{\copy\tempa}\cr}}%
		      \tempc=\wd\tempb
		      \advance\tempc by \wd\tempa
		      \divide\tempc by 2 }
\def\irule#1#2#3{{\irulehelp{#1}{#2}{#3}%
		     \hbox to \wd\tempa{\hss \box\tempb \hss}}}

\begin{document}

\title{Axioms vs. rewrite rules:\\ from completeness to cut elimination}

\author{Gilles Dowek
\thanks{INRIA-Rocquencourt, B.P. 105, 78153 Le Chesnay Cedex,
France. \newline
\texttt{Gilles.Dowek@inria.fr}, 
\texttt{http://coq.inria.fr/\~{}dowek}
}
}

\date{}

\maketitle

\thispagestyle{empty}

\begin{abstract}
Combining a standard proof search method, such as resolution or 
tableaux, and rewriting is a powerful way to cut off search space in
automated theorem proving, but proving the completeness of such
combined methods may be challenging. It may require in particular to
prove cut elimination for an extended notion of
proof that combines deductions and computations. This suggests new
interactions between automated theorem proving and proof theory. 
\end{abstract}

\section*{}

When we search for a proof of the proposition $2 + 2 = 4$, we can use
the axioms of addition and equality to transform this proposition into
$4 = 4$ and 
conclude with the reflexivity axiom.
We could also use these axioms to transform this proposition
into $2 + 2 = 2 + 2$ and conclude, but this proof is redundant with
the 
first, that is in many ways better. Indeed, the axioms of addition are
better used in one direction only: to compute values. In
automated proof  
search systems, we often suppress axioms, such as the axioms of
addition, and replace them by rewrite rules. This permits to cut off
search space while keeping completeness.

The rules of addition apply to terms, but rules applying to
propositions may also be considered. For instance, the axiom 
$$\fa x y~((x \times y) = 0 \Leftrightarrow (x = 0 \vee y = 0))$$
can be replaced by the rule 
$$x \times y = 0 \ra x = 0 \vee y = 0$$
that rewrites an atomic proposition into a disjunction.
Such rules are of special interest in set theory, e.g.
$$x \in \{y,z\} \ra x = y \vee x = z$$
and in type theory \cite{TPM,Holsigma}, e.g. 
$$\varepsilon(x \dot{\Rightarrow} y) \ra \varepsilon (x) \Rightarrow
\varepsilon(y)$$

When we orient an axiom, we want to cut off search space, but we do
not want to loose completeness. In this note, we discuss the
properties that the rewrite system must fulfill, so that orientation
does not jeopardize completeness. 

\section{Orientation}

\subsection{From replacement to rewriting}

We consider a set $A$ and a binary relation $\ra$ defined on $A$. We
write $\ra^{*}$ for its reflexive-transitive closure 
and $\equiv$ for its reflexive-symmetrical-transitive closure.

If $t$ and $t'$ are two elements of $A$, we have $t \equiv t'$ if and
only if there is a sequence of terms $t_{1}, ..., t_{n}$ such  
that 
$t = t_{1}$, $t' = t_{n}$ and 
for each $i$, either $t_{i} \ra t_{i+1}$ or $t_{i} \la t_{i+1}$.

For instance, if $A$ is a set of terms built with a binary operator $+$
and a finite number of constants and $\ra$ is the relation such
that $t \ra u$ if and only if $u$ is obtained from $t$ by replacing a
subterm of the form $(x + y) + z$ by $x + (y + z)$. 
We can establish that 
$$(a + (b + c)) + (d  + e) \equiv (a + b) + ((c + d)  + e)$$
with the sequence

\medskip
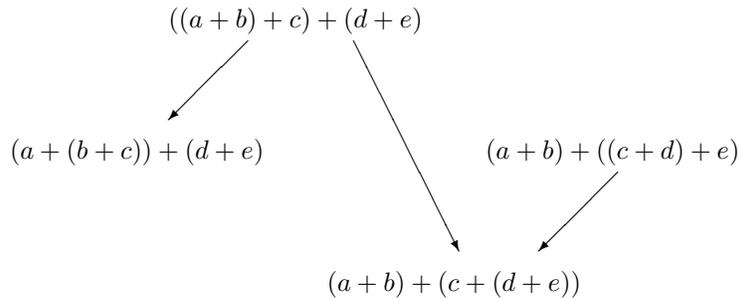
\begin{center}
\begin{picture}(280,100)(0,0)
\put (0,50) {$(a + (b + c)) + (d  + e)$}
\put (60,100) {$((a + b) + c) + (d  + e)$}

\put (90,95) {\vector (-1,-1){30}}
\put (130,95) {\vector (1,-2){40}}
\put (120,0) {$(a + b) + (c + (d  + e))$}
\put (230,45) {\vector (-1,-1){30}}
\put (180,50) {$(a + b) + ((c + d)  + e)$}
\end{picture}
\end{center}

To establish that $t \equiv t'$, we search for such a sequence. 
We can apply the following {\em replacement} method: we start with the
equality $t = t'$ and we derive more equalities with two rules. The
first permits to derive $v = v'$ from $u = u'$ if $u = u'  \ra v = v'$
and the second permits to derive $v = v'$ from $u = u'$ if $u = u' \la
v = v'$. When reach an equality of the form $u = u$, we are done.

Search is more efficient if we restrict to {\em rewriting sequences},
i.e. sequences of the form 
$$t = t_{1} \ra ... \ra t_{n} = u_{1} \la ... \la u_{p} = t'$$
For instance

\medskip
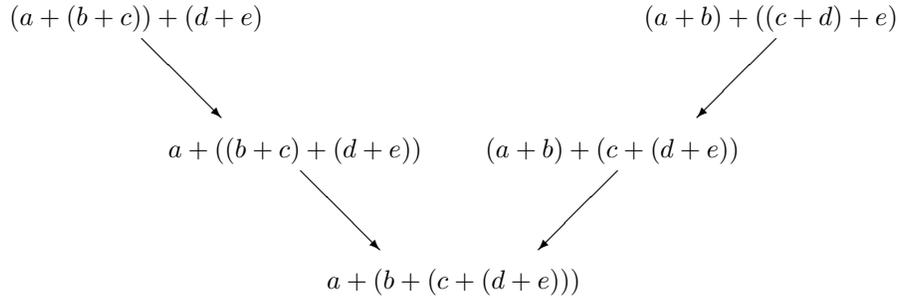
\begin{center}
\begin{picture}(340,100)(0,0)
\put (0,100) {$(a + (b + c)) + (d  + e)$}
\put (50,95) {\vector (1,-1){30}}
\put (60,50) {$a + ((b + c) + (d + e))$}
\put (110,45) {\vector (1,-1){30}}
\put (120,0) {$a + (b + (c + (d + e)))$}
\put (230,45) {\vector (-1,-1){30}}
\put (180,50) {$(a + b) + (c + (d  + e))$}
\put (290,95) {\vector (-1,-1){30}}
\put (240,100) {$(a + b) + ((c + d)  + e)$}
\end{picture}
\end{center}

Indeed, to establish that $t \equiv t'$, we can restrict the
replacement method, to the first rule and rewrite the equality $t =
t'$ to an equality of the form $u = u$. 

Such a restriction may be incomplete. Consider, for
instance, the relation defined by $p \ra' q$ and $p \ra' r$. We have
$q \la' p \ra' r$ but there is no rewriting sequence relating $q$ to
$r$ and the equality $q = r$ cannot be rewritten.

A relation is said to be {\em confluent} if whenever $t \ra^{*} t_{1}$ and 
$t \ra^{*} t_{2}$ there is a object $u$ such that $t_{1} \ra^{*} u$
and $t_{2} \ra^{*} u$.  

For instance, the relation $\ra$ above is confluent, but the relation
$\ra'$ is not. 
It is easy to prove that when a relation is confluent, two objects are
related if and only if they are related by a rewriting sequence and
thus that rewriting is a complete search method.

\subsection{From paramodulation to narrowing}

Let us now turn to proof search in predicate calculus with equality. 
We assume that we have an equality predicate and the associated axioms.
Paramodulation \cite{RobinsonWos} is an extension of resolution that
allows to replace equals by equals in clauses. 

For instance, if we want to refute the clauses 
$$(X + Y) + Z = X + (Y + Z)$$
$$P((a + (b + c)) + (d  + e))$$
$$\neg P((a + b) + ((c + d) + e))$$
we use the associativity to rearrange brackets in the other clauses
until we can resolve them. 

Exactly as above, we can restrict the method and use of some equalities 
in one direction only. We suppress these equalities from the set 
of clauses to refute, we orient them as rewrite rules and we use them
with a {\em narrowing} (or {\em oriented paramodulation}) rule that
permits to unify a subterm of a clause with the left-hand side of a
rewrite rule, and replace the instantiated subterm with the right-hand
side of this rule (see, for instance, \cite{Peterson,HsiangRusi}). 
Notice that since clauses may contain variables, we need to use 
unification (and not matching as in rewriting). 

The completeness of this method mostly rests on confluence of the
rewrite system\footnote{Actually, although confluence plays the major
r\^ole in completeness proofs, termination also is used. Whether or not
completeness fails for non terminating systems in unknown to the
author.}: 
instead of using the rewrite rules in both directions to unfold and fold
terms, we can use them in one direction only and meet on a common
reduct. 

\subsection{Equational resolution}

{\em Equational resolution} \cite{Plotkin,Stickel} is another 
proof search method for predicate calculus with equality. 
In this method, like in narrowing, we suppress some equalities from the
set of clauses to refute and we orient them as rewrite rules. We write
$t \ra t'$ if $t'$ is obtained by rewriting a subterm in $t$. We write
$\ra^{*}$ for the reflexive-transitive closure of this relation and
$\equiv$ for its reflexive-symmetrical-transitive closure. 

Morally, we identify equivalent propositions and we work on equivalence
classes. Hence, we use an extended resolution rules where unification
is replaced  by equational unification \cite{Fay,Hullot}: a unifier of
two-terms $t$ and 
$u$ is a substitution $\sigma$ such that $\sigma t \equiv \sigma u$  
(and not $\sigma t = \sigma u$). For instance the 
terms $(a + (b + c)) + (d  + e)$
and
$(a + b) + ((c + d) + e)$ are trivially unifiable because they are equivalent
and hence with the clauses
$$P((a + (b + c)) + (d  + e))$$
$$\neg P((a + b) + ((c + d) + e))$$
we can apply the resolution rule and get the empty clause.

To solve equational unification problems, we can use narrowing.
But, in contrast with the previous method, the narrowing rule
is applied to the unification problems and not to the clauses.

\section{Completeness}

We focus now on the completeness of equational resolution. We first
recall a completeness proof of resolution and then discuss how it can
be adapted.

\subsection{Completeness of resolution} 
\label{toto1}

\paragraph{Cut elimination} 

A way to prove completeness of resolution is to prove that whenever a
sequent $A_{1}, ..., A_{p} \vdash B_{1}, ..., B_{q}$ has a proof in
sequent calculus (see, for instance,\cite{Gallier,Giraflor}), 
the empty clause can be derived from the clausal form of the propositions
$A_{1}, ..., A_{p}, \neg B_{1}, ..., \neg B_{q}$.

Usually this proof proceeds in two steps: we first prove that if a
sequent
$\Gamma \vdash \Delta$ has a proof, then it has also a proof that does
not use the cut rule of sequent calculus
(cut elimination theorem). 
Then we prove, by induction over proof structure, that if
$\Gamma \vdash \Delta$ has a cut free proof, then the empty clause can
be derived from clausal form $cl(\Gamma, \neg \Delta)$ of the
propositions $\Gamma, \neg \Delta$.  

\paragraph{A variant} 

A variant of this proof isolates the clausification step.
A first lemma proves that if the sequent $\Gamma \vdash \Delta$ has a
proof and $cl(\Gamma, \neg \Delta) = \{P_{1}, ..., P_{n}\}$, then the
sequent $\overline{\fa} P_{1} ... \overline{\fa} P_{n} \vdash$ also
has a proof,
where  $\overline{\fa} P$ is the universal
closure of the proposition $P$.

Then, by the cut elimination theorem, if the sequent
$\overline{\fa} P_{1} ... \overline{\fa} P_{n} \vdash$ has 
a proof, it has also a cut free proof. 
At last, we prove by induction over proof structure that if 
$\overline{\fa} P_{1} ... \overline{\fa} P_{n} \vdash$ has 
a cut free proof then the empty clause can be derived from
$\{P_{1}, ..., P_{n}\} = cl(\Gamma, \neg \Delta)$.

\paragraph{Herbrand theorem}

Instead of using the cut elimination theorem some authors prefer 
to use Herbrand theorem that is a variant of it.

According to Herbrand theorem, if $P_{1}, ..., P_{n}$ are quantifier
free propositions then  
the sequent $\overline{\fa} P_{1} ... \overline{\fa} P_{n} \vdash$ 
has a proof if and only if there
are instances $\sigma_{1}^{1} P_{1}$, ..., $\sigma_{k_{1}}^{1} P_{1}$,
..., $\sigma_{1}^{n} P_{n}$, ..., $\sigma_{k_{n}}^{n} P_{n}$ 
of the propositions 
$P_{1}, ..., P_{n}$ such that the quantifier free proposition 
$\neg (\sigma_{1}^{1} P_{1} \wedge ... \wedge \sigma_{k_{1}}^{1} P_{1}
\wedge ... \wedge \sigma_{1}^{n} P_{n} \wedge ... \wedge
\sigma_{k_{n}}^{n} P_{n})$ 
is tautologous.

As above, a first lemma proves that if the sequent $\Gamma \vdash
\Delta$ has a proof and $cl(\Gamma, \neg \Delta) = \{P_{1}, ...,
P_{n}\}$, then the sequent  
$\overline{\fa} P_{1} ... \overline{\fa} P_{n} \vdash$ also has a proof.
By Herbrand theorem, if the sequent 
$\overline{\fa} P_{1} ... \overline{\fa} P_{n} \vdash$ 
has a proof, then there are instances $\sigma_{1}^{1} P_{1}$, ...,
$\sigma_{k_{1}}^{1} P_{1}$, 
..., $\sigma_{1}^{n} P_{n}$, ..., $\sigma_{k_{n}}^{n} P_{n}$ 
of the propositions 
$P_{1}, ..., P_{n}$ such that the quantifier free proposition 
$\neg (\sigma_{1}^{1} P_{1} \wedge ... \wedge \sigma_{k_{1}}^{1} P_{1}
\wedge ... \wedge \sigma_{1}^{n} P_{n} \wedge ... \wedge
\sigma_{k_{n}}^{n} P_{n})$ 
is tautologous.
At last, we prove that if the proposition
$\neg (\sigma_{1}^{1} P_{1} \wedge ... \wedge \sigma_{k_{1}}^{1} P_{1}
\wedge ... \wedge \sigma_{1}^{n} P_{n} \wedge ... \wedge
\sigma_{k_{n}}^{n} P_{n})$ 
is tautologous then the empty clause can be derived from 
$\{P_{1}, ..., P_{n}\} = cl(\Gamma, \neg \Delta)$.

\subsection{Completeness of equational resolution} 
\label{toto2}

This proof can be adapted to equational resolution. First, the
identification of equivalent propositions used in proof search can be
used in sequent calculus also. This leads to the {\em sequent calculus
modulo} \cite{TPM}. For instance, the axiom rule 
$$\irule{}
        {A \vdash A}
        {\mbox{axiom}}$$
is transformed into the rule
$$\irule{}
        {A \vdash_{\equiv} B}
        {\mbox{axiom if $A \equiv B$}}$$
and the left rule of disjunction 
$$\irule{\Gamma~A \vdash \Delta~~~\Gamma~B \vdash \Delta}
        {\Gamma~A \vee B \vdash \Delta}
        {\mbox{$\vee$-left}}$$
is transformed into the rule
$$\irule{\Gamma~A \vdash_{\equiv} \Delta~~~\Gamma~B \vdash_{\equiv} \Delta}
        {\Gamma~C\vdash_{\equiv} \Delta}
        {\mbox{$\vee$-left if $C \equiv A \vee B$}}$$
In sequent calculus modulo the rules of addition and multiplication,
we have a very short proof that the number $4$ is even
$$\irule{\irule{\irule{}{4 = 4 \vdash_{\equiv} 4 = 2 \times 2}{\mbox{axiom}}}
               {\fa x~(x = x) \vdash_{\equiv} 4 = 2 \times 2}
               {\mbox{$\fa$-left}}
        }
        {\fa x~(x = x) \vdash_{\equiv} \exists y~(4 = 2 \times y)}
        {\mbox{$\exists$-right}}$$
while proving this proposition would be much more cumbersome in 
sequent calculus with the axioms of addition and multiplication. 

Using sequent calculus modulo, we can prove the completeness of
equational resolution. First, we prove the {\em equivalence lemma}: if the
rewrite system encodes a theory ${\cal T}$ then the sequent 
$\Gamma {\cal T} \vdash \Delta$ has a proof in sequent calculus if and
only if the sequent $\Gamma \vdash_{\equiv} \Delta$ has a proof in
sequent calculus modulo. 
Then, the cut elimination theorem extends trivially to sequent
calculus modulo when the equivalence is generated by rewrite rules
applying to terms.  
At last, we prove by induction over proof structure,   
that if $\Gamma \vdash_{\equiv} \Delta$ has a cut free proof in
sequent calculus modulo, then the empty clause can be derived from
$cl(\Gamma, \neg \Delta)$ with the rules of equational resolution.
The confluence of the rewrite system is only needed to prove that 
narrowing is a complete equational unification method.

\subsection{Equational resolution as narrowing}  

Equational resolution can be seen as a formulation of narrowing. 
This suggests another completeness proof, reducing the completeness of
equational resolution to that of narrowing.

Indeed, instead of resolving two clauses and narrowing the unification
problem, as we do in equational unification, we could first narrow the
clauses and apply the usual resolution rule to get the same result.

For instance, instead of resolving 
$$P((a + (b + c)) + (d  + e))$$
$$\neg P((a + b) + ((c + d) + e))$$
and narrowing the unification problem
$$(a + (b + c)) + (d  + e) = (a + b) + ((c + d) + e)$$
we could take an option on these two clauses, and narrow them
until they can be resolved. 

So, narrowing a unification problem is just a way to narrow
the clauses it comes from. Hence, equational resolution can be seen
as a formulation of narrowing and thus as an implementation of 
paramodulation with the restriction that some equations should be used
in one direction only. Hence, the completeness of equational resolution 
rests on confluence.

\section{Rewriting proposition}

So far, we have considered methods where axioms are replaced by
rewrite rules applying to terms. We have seen that the proof search
methods obtained this way could be seen as restrictions of
paramodulation and that their completeness rested on 
confluence.  
We consider now more powerful rewrite rules applying to propositions. 
For instance, the axiom 
$$\fa x y~(x \times y = 0 \Leftrightarrow (x = 0 \vee y = 0))$$
can be replaced by the rule 
$$x \times y = 0 \ra x = 0 \vee y = 0$$
that applies directly to propositions. For technical reasons, we
restrict to rules, such as this one, whose left-hand side is an atomic
proposition. 

For example, with this rule, we can prove the proposition 
$$\exists z~(a \times a = z \Rightarrow a = z)$$
in sequent calculus modulo
$$\irule{\irule{\irule{\irule{}
                             {a = 0 \vdash_{\equiv} a = 0}
                             {\mbox{axiom}}
                       ~~~~~~~~~~~~~~~
                       \irule{}
                             {a = 0 \vdash_{\equiv} a = 0}
                             {\mbox{axiom}}
                      }
                      {a \times a = 0 \vdash_{\equiv} a = 0}
                      {\mbox{$\vee$-left}}
               }
               {\vdash_{\equiv} a \times a = 0 \Rightarrow a = 0}
               {\mbox{$\Rightarrow$-right}}
               }
        {\vdash_{\equiv} \exists z~(a \times a = z \Rightarrow a = z)}
        {\mbox{$\exists$-right}}$$

\subsection{Resolution modulo}

In this case, equational resolution can be extended to {\em resolution
modulo} \cite{TPM}. In resolution modulo, like in equational
resolution, the rewrite rules applying to terms are used to narrow
unification problems, but, like in narrowing, the rules applying to
propositions are used to narrow the clauses directly.

For instance, if we take the clausal form of the negation of the
proposition  
$$\exists z~(a \times a = z \Rightarrow a = z)$$
i.e. the clauses
$$a \times a = Z$$
$$\neg a = Z$$
we can narrow the first clause with the rule 
$$x \times y = 0 \ra x = 0 \vee y = 0$$
yielding the clause
$$a = 0$$
Then, we resolve this clause with $\neg a = Z$ and get the empty
clause.

\subsection{Resolution with axioms} 

An alternative is to keep the axiom 
$$\fa x y~(x \times y = 0 \Leftrightarrow (x = 0 \vee y = 0))$$ 
and to use resolution. In this case, we have to refute the clauses
$$\neg X \times Y = 0, X = 0, Y = 0$$
$$X \times Y = 0, \neg X = 0$$
$$X \times Y = 0, \neg Y = 0$$
$$a \times a = Z$$
$$\neg a = Z$$

We resolve the clause $a \times a = Z$ with the clause $\neg X \times
Y = 0, X = 0, Y = 0$ yielding the clause $a = 0$.
Then, we resolve this clause with $\neg a = Z$ and get the empty
clause.

An above, resolution modulo can be seen as a restriction of
resolution, where some clauses can be used in one direction only. As
above we could think that resolution modulo is complete as soon as the
rewrite system is confluent. Unfortunately this is not the case. 

\subsection{A counter example to completeness}

The axiom
$$A \Leftrightarrow (B \wedge \neg A)$$
can be transformed into a rewrite rule (Crabb\'e's rule)
$$A \ra B \wedge \neg A$$
Resolution modulo cannot prove the proposition $\neg B$, because from
the clause $B$, neither the resolution rule nor the narrowing rule can
be applied.

But, surprisingly, with the axiom $A \Leftrightarrow (B \wedge \neg
A)$, we can prove the proposition $\neg B$.
Indeed, the clausal form of the proposition 
$A \Rightarrow (B \wedge \neg A)$ yields the clauses
$$\neg A, B~~~~~\neg A$$
the clausal form of the proposition 
$(B \wedge \neg A) \Rightarrow A$ yields
$$A, \neg B$$
and the clausal form of the negation of the proposition $\neg B$ yields
$$B$$
From $B$ and $A, \neg B$, we derive $A$ and then from this clause and 
$\neg A$ we get the empty clause.
Hence, a resolution proof with axioms cannot always be transformed into
a resolution modulo proof. Resolution modulo cannot be seen as a
restriction of resolution where some clauses can be used in one
direction only and completeness may be lost even if the rewrite system
is confluent.

Notice that, in the resolution proof, clauses coming from both propositions 
$A \Rightarrow (B \wedge \neg A)$ and
$(B \wedge \neg A) \Rightarrow A$
are used. 
Hence, although the rewrite system is confluent, we need both to
unfold $A$ to $B \wedge \neg A$ and to fold 
$B \wedge \neg A$ to $A$. 
Moreover, when we resolve $B$ with $\neg B, A$ we fold 
$B \wedge \neg A$ to $A$ although we have proved the proposition 
$B$, but not the proposition $\neg A$ yet. This {\em partial folding}
that resolution allows but  
resolution modulo disallows is the reason why some
resolution proofs cannot be transformed into resolution modulo proofs. 

Hence, orientation cuts off search space more dramatically when we
have rules rewriting propositions that when we have only rules
rewriting terms. It can cut off search space so dramatically that
completeness may be lost.

\subsection{Completeness of resolution modulo}

Since resolution can prove the proposition $\neg B$ with the axiom 
$A \Leftrightarrow (B \wedge \neg A)$, the sequent $A \Leftrightarrow
(B \wedge \neg A) \vdash \neg B$ can be proved in sequent calculus and
hence the sequent $\vdash \neg B$ can be proved in sequent calculus
modulo. For instance, it has the proof 
$$\irule{\irule{\irule{\irule {}
                              {B \vdash B}
                              {\mbox{axiom}}
                     ~~~~~~~~~~~~~
                     \irule{\irule{\irule{\irule{\irule{}
                                                       {B, A, B \vdash A}
                                                       {\mbox{weak. + ax.}}
                                                }
                                                {B, \neg A, A, B \vdash}
                                                {\mbox{$\neg$-left}}
                                         }
                                         {A, A, B \vdash}
                                         {\mbox{$\wedge$-left}}
                                  }
                                  {A, B \vdash}
                                  {\mbox{contraction-left}}
                           }
                           {B \vdash \neg A}
                           {\mbox{$\neg$-right}}
                    }
                    {B \vdash A}
                    {\mbox{$\wedge$-right}}
              ~~~~~~~~~~~~~~~~~~~~~~~~~~~~~~~~~~~~~~~~
              \irule{\irule{\irule{\irule{}
                                         {B, A, B \vdash A}
                                         {\mbox{weak. + ax.}}
                                  }
                                  {B, \neg A, A, B \vdash}
                                  {\mbox{$\neg$-left}} 
                           }
                           {A, A, B \vdash}
                           {\mbox{$\wedge$-left}}
                    }
                    {A, B \vdash }
                    {\mbox{contraction-left}}
             }
             {B \vdash}
             {\mbox{cut}}  
      }
      {\vdash \neg B}
      {\mbox{$\neg$-right}}
$$
The proposition $\neg B$ has a proof in sequent calculus modulo
but not in resolution modulo and thus resolution modulo is incomplete. 

The completeness theorem of resolution modulo does not prove that each
time a proposition has a proof in sequent calculus modulo, it has a
proof in resolution modulo (because this is false), but it proves, 
using techniques similar to those developed in section
\ref{toto1} and \ref{toto2}, 
that if
a proposition has a cut free proof in sequent calculus modulo, it has
a proof in resolution modulo.

The proposition $\neg B$ has proof in sequent calculus modulo, but no
cut free proof and sequent calculus modulo the rule $A \ra B \wedge
\neg A$ does not have the cut elimination property. 

Modulo some rewrite systems such as 
$$x \times y = 0 \ra x = 0 \vee y = 0$$
or
$$A \ra B \wedge A$$
sequent calculus modulo has the cut elimination property and thus
resolution modulo is complete. Hence modulo these rewrite systems 
there is no occurrence of the phenomenon we have observed above, where
a resolution proof could not be transformed into a resolution modulo
proof because it used partial folding. Still the translation of
resolution proofs to resolutions modulo proofs is non trivial because
it involves cut elimination. 

In \cite{DowekWerner} we present cut elimination proofs for 
large classes of rewrite systems 
(including various presentations of type theory, all confluent and
terminating quantifier free rewrite systems, ...) and we conjecture
that cut elimination holds for all terminating and confluent rewrite
system, although we know that such a conjecture cannot be proved in
type theory or in large fragments of set theory because it implies
their consistency. 
Cut elimination also holds for
sequent calculus modulo some non terminating rewrite systems such as 
$$A \ra B \wedge A$$

Notice that when resolution modulo is complete (for instance for type
theory), this completeness result cannot be proved in the theory
itself (because it implies its consistency), while
confluence usually can. So, it is not surprising that tools more
powerful than confluence, such as cut elimination, are required. 

\subsection{Towards a new kind of completion}

Completion \cite{KB} transforms a rewrite system into a
confluent one. We can imagine a similar process that transforms  
a rewrite system into one modulo which sequent calculus has the cut
elimination property.

For instance, the rule $A \ra B \wedge \neg A$ is equivalent to the axiom 
$A \Leftrightarrow (B \wedge \neg A)$ whose clausal form is 
$$A, \neg B~~~~~B, \neg A~~~~~\neg A$$
like that of the axioms
$A \Leftrightarrow \bot$
and 
$B \Leftrightarrow A$ 
that are equivalent to the rewrite system
$A \ra \bot$, $B \ra A$ modulo which sequent calculus has the cut
elimination property and resolution is complete. 

So we could imagine to transform the rewrite system
$$A \ra (B \wedge \neg A)$$
into 
$$A \ra \bot$$
$$B \ra A$$
or even into
$$A \ra \bot$$
$$B \ra \bot$$
in order to recover completeness.

\section*{Acknowledgements}
Thanks to Th.~Hardin, C.~Kirchner, Ch.~Lynch and B.~Werner for many
helpful discussions on this subject.

\end{document}